\documentclass[runningheads]{llncs}
\usepackage[T1]{fontenc}
\usepackage{amsfonts}
\usepackage{graphicx}
\usepackage{amsmath}
\usepackage{amssymb}
\usepackage{algorithm}
\usepackage{algpseudocode}
\usepackage{booktabs}
\usepackage{multirow}
\usepackage{tikz}
\usetikzlibrary{shapes.geometric, arrows.meta}

\begin{document}
\title{TailSpec-EASE: Knowledge-Graph-Regularized Linear Recommendation for Web Long-Tail Discovery}

\titlerunning{TailSpec-EASE: KG-Regularized Linear Recommendation}

\author{Jianru Shen\orcidID{0009-0000-3546-9616}}
\authorrunning{J. Shen}
\institute{University of Montana, Missoula, MT 59812, USA \\
\email{js258133@umconnect.umt.edu}}
\maketitle              
\begin{abstract}
Recommender systems on Web platforms tend to over-serve popular items and
neglect the long tail. Item-side knowledge graphs (KGs), often available as
linked data or RDF-style Web resources, can help by connecting sparse items
through shared semantic attributes. Many competitive KG-aware recommenders
rely on graph neural architectures, whereas strong shallow linear models such as
EASE$^\text{R}$ typically ignore side information and may become infeasible in
their global closed-form version. We introduce TailSpec-EASE, a lightweight
recommender that injects a relation-aware spectral KG prior into a local
closed-form reconstruction objective. The prior strength adapts to item
popularity, giving stronger semantic guidance to long-tail items.

Across four public benchmarks and a broad set of classical, linear, graph-CF,
KG-aware neural, and score-level KG baselines, TailSpec-EASE attains a favorable
trade-off between overall accuracy, long-tail performance, and training cost.
It improves NDCG@20 by up to $24\%$ over its counterpart without KG
information. All tail-metric improvements over the no-KG counterpart are
statistically significant under a paired bootstrap, and overall NDCG improves
significantly on three of the four datasets. In a representative Amazon-book
timing study, TailSpec-EASE trains in $37$ seconds on CPU, compared with
$2{,}584$ seconds for a GPU-trained KGAT run and $15{,}800$ seconds for CPU
LightGCN, while attaining higher NDCG@20 and Tail Recall@20 on that dataset.
It also remains feasible on catalogs where the global closed-form model runs out
of memory.

\end{abstract}

\section{Introduction}
\label{sec:intro}

Recommendation is a core component of modern Web platforms, where users must
discover relevant items from large catalogs. A persistent challenge in such
Web-scale settings is popularity bias: a small set of head items receives most
exposure, while niche long-tail items are rarely recommended despite their value
to users and providers~\cite{park2008longtail,steck2011popularity}. Item-side
knowledge graphs (KGs), increasingly published on the Web as linked open data
and RDF-style resources, can alleviate this problem by connecting sparse items
through shared semantic attributes such as genres, authors, or categories
~\cite{wang2019kgat,wang2018ripplenet}. Exploiting such structured Web
knowledge while keeping recommendation efficient and scalable is a central
concern for Web information systems.

Existing methods leave a practical gap. A major line of KG-aware recommenders
uses graph neural architectures to learn or propagate KG representations
end-to-end~\cite{wang2019kgat,wang2019kgcn}. These models are expressive but
often require GPU training, long optimization schedules, and careful tuning,
which limits reproducibility and deployment in commodity Web infrastructure.
In parallel, shallow linear recommenders such as SLIM and EASE$^\text{R}$ are
simple, fast, and highly competitive~\cite{ning2011slim,steck2019ease,dacrema2019recsys},
but they ignore side information, and the global closed-form solution of
EASE$^\text{R}$ requires a dense item--item inversion that becomes
memory-prohibitive on large catalogs. This raises a practical question: can
item-side KG structure be incorporated into a lightweight local linear
recommender that remains scalable on Web-scale catalogs while improving
long-tail exposure?

We address this question with TailSpec-EASE. The method distills typed KG
relations into a sparse spectral prior over items, injects this prior directly
into a local closed-form reconstruction objective, and adapts the KG strength to
item popularity so that tail items receive stronger semantic guidance. Across
four public benchmarks, TailSpec-EASE improves long-tail performance while
remaining competitive in overall accuracy. A central finding is that the point
of KG injection matters: model-level regularization consistently outperforms
applying the same KG signal as score-level post-processing.

\medskip
\noindent\textbf{Contributions.}
Guided by the Web-catalog setting above, this paper makes four contributions.
\textbf{(1)} We propose TailSpec-EASE, a closed-form local linear recommender
that injects a relation-aware KG prior into item reconstruction and avoids dense
global item--item inversion, making it suitable for large Web catalogs.
\textbf{(2)} We introduce popularity-adaptive KG regularization, which gives
stronger semantic guidance to long-tail items while allowing collaborative
evidence to dominate for head items.
\textbf{(3)} We provide a controlled comparison of model-level and score-level
KG injection, showing that objective-level injection gives a better
accuracy--long-tail trade-off than post-processing the scores.
\textbf{(4)} We evaluate against classical, linear, graph-CF, KG-aware neural,
and score-level KG baselines, showing a favorable accuracy--tail--efficiency
trade-off: TailSpec-EASE remains competitive with KGAT in overall NDCG,
improves tail recall over KGAT on all four datasets, and trains in $37$ seconds
on CPU in a representative Amazon-book timing study.

\section{Related Work}
\label{sec:related}

\paragraph{Knowledge-graph-aware recommendation.}
KG-aware recommenders enrich collaborative filtering with item-side structured
knowledge, much of it drawn from Web knowledge bases and linked open data.
Embedding-based methods such as RippleNet propagate user preferences over the
KG~\cite{wang2018ripplenet}, while graph neural methods such as KGCN,
KGNN-LS, and KGAT aggregate multi-hop KG neighborhoods
~\cite{wang2019kgcn,wang2019kgnnls,wang2019kgat}. These methods are expressive
but often require end-to-end representation learning, GPU training, and careful
hyperparameter tuning. TailSpec-EASE takes a complementary approach: it uses
the KG as a fixed relation-aware prior and injects this prior into a closed-form
linear objective, avoiding learned KG embeddings entirely.

\paragraph{Linear, long-tail, and post-processing recommenders.}
Shallow linear recommenders remain strong baselines. SLIM learns sparse
item--item weights by regularized regression~\cite{ning2011slim}, and
EASE$^\text{R}$ gives a competitive closed-form $\ell_2$-regularized variant
~\cite{steck2019ease}; reproducibility studies show that such models often match
or exceed more complex neural recommenders~\cite{dacrema2019recsys}. Most such
models operate only on the interaction matrix. Separately, long-tail and
popularity-aware methods often rely on re-ranking, popularity control, or
calibrated post-processing~\cite{abdollahpouri2017controlling,steck2018calibrated,liu2020longtail}.
TailSpec-EASE differs from both lines: it addresses the long tail inside the
model objective through a KG-derived prior whose strength adapts to item
popularity, and Sect.~\ref{sec:postproc} shows that this model-level design
outperforms score-level KG post-processing.

\section{The TailSpec-EASE Method}
\label{sec:method}

\begin{figure}[t]
\centering
\includegraphics[width=\textwidth]{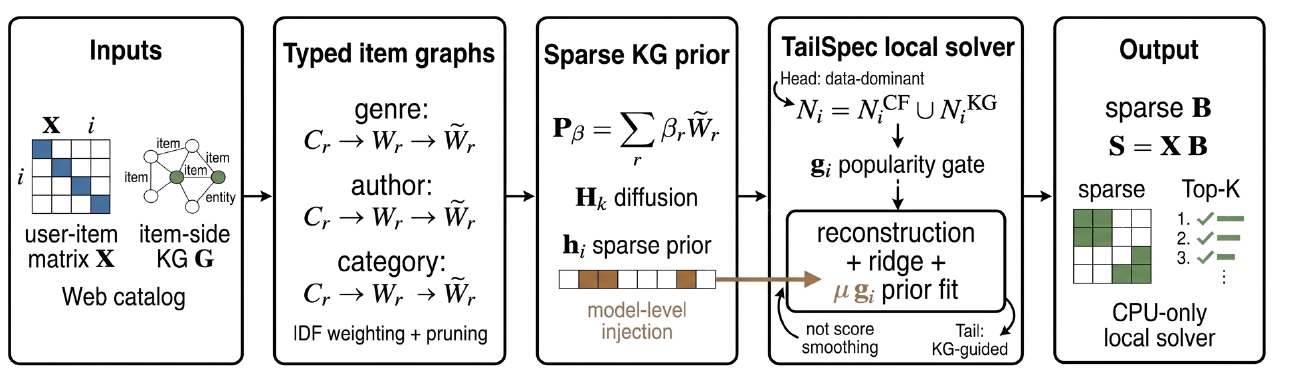}
\caption{Overview of TailSpec-EASE. The item-side KG is converted into typed
item--item graphs, distilled into a sparse relation-aware prior, and injected
directly into the local reconstruction objective rather than applied as
score-level post-processing. The popularity-adaptive gate $g_i$ gives stronger
KG guidance to long-tail items, while the local closed-form solver avoids dense
global item--item inversion.}
\label{fig:overview}
\end{figure}

TailSpec-EASE is a lightweight item-based recommender that injects a
relation-aware knowledge-graph (KG) prior directly into a local linear
reconstruction objective, with a regularization strength that adapts to item
popularity. As shown in Fig.~\ref{fig:overview}, item-side KG triples are
converted into typed item--item graphs, distilled into a sparse spectral prior,
and injected into a local closed-form solver rather than applied as score-level
post-processing.

\subsection{Problem Setting and Notation}
\label{sec:method:setup}

Let $\mathcal{U}=\{u_1,\dots,u_m\}$ be a set of $m$ users and
$\mathcal{I}=\{i_1,\dots,i_n\}$ a set of $n$ items. User feedback is implicit
and encoded by the binary interaction matrix
$\mathbf{X}\in\{0,1\}^{m\times n}$, where $\mathbf{X}_{ui}=1$ if user $u$ has
interacted with item $i$ and $\mathbf{X}_{ui}=0$ otherwise. We write
$\mathbf{x}_i\in\{0,1\}^{m}$ for the $i$-th column of $\mathbf{X}$, i.e. the
interaction vector of item $i$ over all users.

Item-side structured knowledge is given by a knowledge graph
$\mathcal{G}=(\mathcal{E},\mathcal{R},\mathcal{T})$, where $\mathcal{E}$ is a
set of entities with $\mathcal{I}\subseteq\mathcal{E}$ (every item is an
entity), $\mathcal{R}$ is a set of relation types, and
$\mathcal{T}\subseteq\mathcal{E}\times\mathcal{R}\times\mathcal{E}$ is a set of
triples. A triple $(h,r,t)\in\mathcal{T}$ states that head entity $h$ is linked
to tail entity $t$ under relation $r$. Examples of relations include
\emph{film--genre}, \emph{book--author}, or \emph{business--category}.

Item-based linear recommenders predict scores as $\mathbf{S}=\mathbf{X}\mathbf{B}$,
where $\mathbf{B}\in\mathbb{R}^{n\times n}$ is an item--item weight matrix and
$\mathbf{S}\in\mathbb{R}^{m\times n}$ collects the predicted user--item scores.
EASE$^\text{R}$~\cite{steck2019ease} learns $\mathbf{B}$ in closed form by
minimizing $\lVert\mathbf{X}-\mathbf{X}\mathbf{B}\rVert_F^2
+\lambda\lVert\mathbf{B}\rVert_F^2$ subject to
$\mathrm{diag}(\mathbf{B})=\mathbf{0}$, which requires inverting the dense
$n\times n$ Gram matrix and becomes memory-prohibitive for large catalogs.
SLIM~\cite{ning2011slim} instead learns each column of $\mathbf{B}$
independently as a regularized regression of one item on the others. Our method
follows the local, per-item formulation of SLIM but retains a ridge closed-form
solution, and augments it with a KG prior; we therefore avoid any dense
$n\times n$ inversion.

Given $\mathbf{X}$ and $\mathcal{G}$, our goal is to learn a sparse
$\mathbf{B}$ that produces a high-quality ranking of unseen items for each user,
with particular emphasis on \emph{long-tail} items, i.e. items with few training
interactions. We denote the popularity of item $i$ by
$d_i=\sum_{u\in\mathcal{U}}\mathbf{X}_{ui}$, its number of observed
interactions.

\subsection{Relation-Aware Item Graphs}
\label{sec:method:graphs}

The first stage turns the KG into a set of item--item graphs, one per relation
type, so that semantic proximity between items can later inform the
reconstruction weights.

\paragraph{Per-relation incidence.}
For each relation $r\in\mathcal{R}$ we build an item--entity incidence matrix
$\mathbf{C}_r\in\{0,1\}^{n\times|\mathcal{E}|}$, where
$(\mathbf{C}_r)_{ie}=1$ if item $i$ is linked to entity $e$ by a triple under
relation $r$ in either direction, and $0$ otherwise. Thus $\mathbf{C}_r$
records which entities each item is attached to through relation $r$.

\paragraph{Inverse-document-frequency weighting.}
Entities differ greatly in degree: broad attributes connect many items and are
less discriminative than specific entities. We therefore down-weight entity $e$
by its inverse document frequency,
\begin{equation}
\mathrm{idf}(e)=\log\frac{n}{\mathrm{df}(e)},
\qquad
\mathrm{df}(e)=\sum_{i\in\mathcal{I}}(\mathbf{C}_r)_{ie},
\label{eq:idf}
\end{equation}
where $\mathrm{df}(e)$ is the number of items attached to entity $e$. Entities
with extremely large degree (above a cutoff $\tau$) are removed, both to
suppress noise and to keep the construction tractable for large catalogs.

\paragraph{Typed item--item graph.}
The weighted item--item adjacency for relation $r$ is
\begin{equation}
\mathbf{W}_r=\mathbf{C}_r\,\mathbf{D}^{\mathrm{idf}}\,\mathbf{C}_r^{\top},
\qquad
\mathbf{D}^{\mathrm{idf}}=\mathrm{diag}\!\big(\mathrm{idf}(e)\big)_{e\in\mathcal{E}},
\label{eq:Wr}
\end{equation}
so that $(\mathbf{W}_r)_{ij}$ accumulates the idf-weighted entities shared by
items $i$ and $j$ under relation $r$. We set the diagonal to zero and retain
only the $m_W$ largest entries per row to preserve sparsity. Each
$\mathbf{W}_r$ is then symmetrically normalized,
\begin{equation}
\widetilde{\mathbf{W}}_r
=\mathbf{D}_r^{-1/2}\,\mathbf{W}_r\,\mathbf{D}_r^{-1/2},
\qquad
\mathbf{D}_r=\mathrm{diag}\!\Big(\textstyle\sum_j(\mathbf{W}_r)_{ij}\Big),
\label{eq:Wnorm}
\end{equation}
where $\mathbf{D}_r$ is the diagonal degree matrix of $\mathbf{W}_r$. The
normalization prevents high-degree items from dominating the subsequent
propagation~\cite{chung1997spectral}.

\paragraph{Relation-aware operator.}
The normalized typed graphs are combined into a single operator
\begin{equation}
\mathbf{P}_{\boldsymbol{\beta}}=\sum_{r\in\mathcal{R}}\beta_r\,
\widetilde{\mathbf{W}}_r,
\qquad
\beta_r\ge 0,\quad\sum_{r\in\mathcal{R}}\beta_r=1,
\label{eq:Pbeta}
\end{equation}
where the weights $\boldsymbol{\beta}=(\beta_r)_{r\in\mathcal{R}}$ control the
contribution of each relation type and are selected on held-out data. Setting
all $\beta_r$ equal recovers an untyped combination, which we use as an ablation
in Sect.~\ref{sec:ablation}.

\subsection{Spectral Knowledge-Graph Prior}
\label{sec:method:prior}

The operator $\mathbf{P}_{\boldsymbol{\beta}}$ defines one-hop semantic
proximity. To capture multi-hop structure while controlling smoothing, we apply
a polynomial graph filter. For a non-negative integer order $K$ and
coefficients $c_0,\dots,c_K$, the filter is
\begin{equation}
\mathbf{H}_K=\sum_{k=0}^{K}c_k\,\mathbf{P}_{\boldsymbol{\beta}}^{\,k}.
\label{eq:filter}
\end{equation}
Because $\mathbf{H}_K$ is a polynomial in a normalized graph operator, it is a
\emph{spectral graph filter}: it acts on the eigenvalues of
$\mathbf{P}_{\boldsymbol{\beta}}$ without an explicit eigendecomposition. In
practice we use a restart-weighted form, $c_k=(1-\rho)\rho^{k}$ with decay
$\rho\in[0,1)$, which yields the truncated personalized-PageRank
diffusion~\cite{page1999pagerank,gori2007itemrank}
\begin{equation}
\mathbf{H}_K=(1-\rho)\sum_{k=0}^{K}\rho^{k}\,
\mathbf{P}_{\boldsymbol{\beta}}^{\,k}.
\label{eq:ppr}
\end{equation}
The term $k=0$ preserves an item's own identity, $k=1$ uses direct KG
neighbors, and larger $k$ reaches farther but risks over-smoothing; we study the
effect of $K$ in Sect.~\ref{sec:ablation}.

To keep the prior sparse and well-scaled, we compute it row by row. For item
$i$ we propagate the indicator vector $\mathbf{e}_i$ through
$\mathbf{P}_{\boldsymbol{\beta}}$ up to $K$ times, truncating to the largest
entries after each multiplication so that no dense intermediate ever forms. The
resulting row is restricted to its $m_H$ largest entries, the self-entry is
removed, and the row is normalized to sum to one:
\begin{equation}
\mathbf{h}_i=\frac{\hat{\mathbf{h}}_i}{\lVert\hat{\mathbf{h}}_i\rVert_1},
\qquad
\hat{\mathbf{h}}_i=\mathrm{top}\text{-}m_H\big((\mathbf{H}_K)_{i,:}\big),
\quad(\mathbf{h}_i)_i=0,
\label{eq:hi}
\end{equation}
where $(\mathbf{H}_K)_{i,:}$ is the $i$-th row of $\mathbf{H}_K$ and
$\mathrm{top}\text{-}m_H(\cdot)$ keeps its $m_H$ largest entries. The
$\ell_1$-normalization makes $\mathbf{h}_i$ a distribution over KG-related
items, which fixes the scale of the prior and lets a single hyperparameter
control its influence in the objective below. Items with no KG links yield an
empty $\mathbf{h}_i$ and fall back to a purely collaborative reconstruction.

\subsection{Tail-Adaptive Regularization}
\label{sec:method:tail}

The KG prior is most valuable for items with weak collaborative evidence. We
therefore scale its strength per item by a gate that decreases with popularity,
\begin{equation}
g_i=\frac{1}{\big(1+\log(1+d_i)\big)^{\gamma}},
\qquad\gamma\ge 0,
\label{eq:gate}
\end{equation}
where $d_i$ is the training popularity of item $i$ and $\gamma$ controls how
sharply the gate distinguishes head from tail items. Long-tail items have small
$d_i$ and thus a gate value $g_i$ close to one, while popular items have large
$d_i$ and a small $g_i$. The case $\gamma=0$ gives $g_i\equiv 1$, a constant
gate that applies the prior uniformly; we use it as an ablation in
Sect.~\ref{sec:ablation}.

\subsection{Local Objective and Closed-Form Solver}
\label{sec:method:solver}

We avoid learning a dense $\mathbf{B}$ by solving an independent local problem
for each item over a small candidate neighborhood.

\paragraph{Hybrid neighborhood.}
For target item $i$ we form the candidate set
\begin{equation}
\mathcal{N}_i=\mathcal{N}_i^{\mathrm{CF}}\cup\mathcal{N}_i^{\mathrm{KG}},
\qquad i\notin\mathcal{N}_i,
\label{eq:nbr}
\end{equation}
where $\mathcal{N}_i^{\mathrm{CF}}$ contains the $m_{\mathrm{CF}}$ items most
similar to $i$ by cosine similarity on the interaction matrix, and
$\mathcal{N}_i^{\mathrm{KG}}$ contains the items with nonzero entries in the KG
prior $\mathbf{h}_i$. Collaborative neighbors preserve behavioral relevance,
while KG neighbors add semantic support that is especially useful when
interactions are sparse. We write $\mathbf{X}_{\mathcal{N}_i}\in\{0,1\}^{m\times
|\mathcal{N}_i|}$ for the submatrix of $\mathbf{X}$ with the columns in
$\mathcal{N}_i$, and $\mathbf{h}_i^{\mathrm{loc}}\in\mathbb{R}^{|\mathcal{N}_i|}$
for the prior $\mathbf{h}_i$ restricted to those columns.

\paragraph{Objective.}
Let $\mathbf{b}_i\in\mathbb{R}^{|\mathcal{N}_i|}$ be the reconstruction weights
of item $i$ over its neighborhood. We minimize
\begin{equation}
\min_{\mathbf{b}_i}\;
\underbrace{\big\lVert\mathbf{x}_i-\mathbf{X}_{\mathcal{N}_i}\mathbf{b}_i
\big\rVert_2^2}_{\text{collaborative reconstruction}}
+\underbrace{\lambda\lVert\mathbf{b}_i\rVert_2^2}_{\text{ridge}}
+\underbrace{\mu\,g_i\big\lVert\mathbf{b}_i-\mathbf{h}_i^{\mathrm{loc}}
\big\rVert_2^2}_{\text{tail-adaptive KG prior}},
\label{eq:obj}
\end{equation}
where $\lambda\ge 0$ controls the ridge penalty and $\mu\ge 0$ controls the
overall strength of the KG prior. The first term fits the observed
interactions, the second stabilizes the weights, and the third pulls
$\mathbf{b}_i$ toward the KG prior with a strength $\mu g_i$ that is larger for
tail items. Setting $\mu=0$ recovers a purely collaborative local model.

\paragraph{Closed-form solution.}
Equation~\eqref{eq:obj} is a ridge regression with a non-zero prior mean and has
the closed-form solution
\begin{equation}
\mathbf{b}_i=\big(\mathbf{X}_{\mathcal{N}_i}^{\top}\mathbf{X}_{\mathcal{N}_i}
+(\lambda+\mu g_i)\mathbf{I}\big)^{-1}
\big(\mathbf{X}_{\mathcal{N}_i}^{\top}\mathbf{x}_i+\mu g_i\,
\mathbf{h}_i^{\mathrm{loc}}\big),
\label{eq:solve}
\end{equation}
where $\mathbf{I}$ is the $|\mathcal{N}_i|\times|\mathcal{N}_i|$ identity matrix.
The local Gram block
$\mathbf{X}_{\mathcal{N}_i}^{\top}\mathbf{X}_{\mathcal{N}_i}$ is assembled only
for the selected neighborhood. In implementation we cache sparse, pruned
item--item co-occurrence information and index the entries needed for each local
block, rather than materializing a dense $n\times n$ Gram matrix.

\paragraph{Scoring and complexity.}
The solved weights are placed into column $i$ of a sparse matrix
$\mathbf{B}\in\mathbb{R}^{n\times n}$, with support $\mathcal{N}_i$. Final
scores are $\mathbf{S}=\mathbf{X}\mathbf{B}$, and items already seen by a user
are masked before ranking. Each subproblem solves a linear system of size
$|\mathcal{N}_i|\le m_{\mathrm{CF}}+m_H$, so the cost is governed by the chosen
neighborhood size rather than the catalog size $n$. This is the key difference
from global EASE$^\text{R}$: we never form or invert the dense $n\times n$
matrix, which is what makes the method feasible on commodity hardware for large
catalogs.

\section{Experimental Setup}
\label{sec:setup}

\subsection{Datasets}
\label{sec:setup:data}

We evaluate on four public benchmarks with item-side knowledge graphs that span
two orders of magnitude in interaction density: MovieLens-1M
(ML-1M)~\cite{harper2015movielens}, Amazon-book~\cite{mcauley2015image},
Last-FM (Last.fm, HetRec~2011)~\cite{cantador2011hetrec}, and the Yelp Open
Dataset (Yelp2018)~\cite{yelp2018dataset}. We use the KG-augmented versions
released with prior work: Amazon-book, Last-FM, and Yelp2018 follow the standard
KGAT splits and Freebase-derived KGs~\cite{wang2019kgat}, while ML-1M uses the
RippleNet KG~\cite{wang2018ripplenet}, with ratings of at least $4$ converted to
implicit positives, only KG-linked items retained, and iterative ten-core
filtering, keeping preprocessing consistent across datasets.
Table~\ref{tab:datasets} summarizes the resulting statistics. Popularity groups
are computed from training interactions only: the top $20\%$ of items form the
head, the next $30\%$ the middle, and the bottom $50\%$ the tail.

\begin{table}[t]
\centering
\caption{Dataset statistics. Head/middle/tail are item counts by training
popularity (top $20\%$ / next $30\%$ / bottom $50\%$). Interactions are the
combined train and test positives.}
\label{tab:datasets}
\setlength{\tabcolsep}{4pt}
\resizebox{\textwidth}{!}{%
\begin{tabular}{lrrrrrrrrr}
\toprule
Dataset & Users & Items & Interac. & Density & Rel. & Triples & Head & Mid & Tail\\
\midrule
ML-1M       & 5{,}648  & 2{,}249  & 374{,}142   & 2.945\% & 12 & 1{,}241{,}995 & 449   & 674    & 1{,}126\\
Amazon-book & 70{,}679 & 24{,}915 & 846{,}434   & 0.048\% & 39 & 2{,}557{,}746 & 4{,}983 & 7{,}474 & 12{,}458\\
Last-FM     & 23{,}566 & 48{,}123 & 3{,}034{,}763 & 0.268\% & 9  & 464{,}567   & 9{,}624 & 14{,}436 & 24{,}063\\
Yelp2018    & 45{,}919 & 45{,}538 & 1{,}183{,}610 & 0.057\% & 42 & 1{,}853{,}704 & 9{,}107 & 13{,}661 & 22{,}770\\
\bottomrule
\end{tabular}%
}
\end{table}

\subsection{Evaluation Protocol and Metrics}
\label{sec:setup:protocol}

We adopt a leakage-controlled protocol. For each dataset we use the standard
test split and hold out a further $10\%$ of each user's training interactions as
validation. Hyperparameters are tuned only on validation, item popularity is
computed only from training interactions, and the test set is evaluated once
with the validation-selected configuration. For every user we rank all items not
seen in training and report Recall@20 and NDCG@20~\cite{jarvelin2002ndcg}. For
long-tail behavior, we also report the same metrics restricted to each
popularity group, where the group-restricted score considers only ground-truth
items in that group. Unless stated otherwise,
significance is assessed by paired user-level bootstrap with $1{,}000$ user
resamples; a $95\%$ confidence interval excluding zero indicates significance.

\subsection{Baselines}
\label{sec:setup:baselines}

We compare against five families of baselines: classical recommenders
(Popularity, ItemKNN~\cite{sarwar2001itemknn}, iALS~\cite{hu2008ials}, and
BPR-MF~\cite{rendle2009bpr}); shallow linear recommenders (global
EASE$^\text{R}$~\cite{steck2019ease} and Local EASE, Eq.~\eqref{eq:obj} with
$\mu=0$); a graph collaborative-filtering reference
(LightGCN~\cite{he2020lightgcn}); a KG-aware neural recommender
(KGAT~\cite{wang2019kgat}); and two score-level KG post-processing baselines.
Smooth applies one-hop propagation $\mathbf{S}\mathbf{P}_{\boldsymbol{\beta}}$ to
Local EASE scores, and Diffuse applies the multi-hop prior
$\mathbf{S}\mathbf{H}_K$, each with a validation-tuned mixing weight. KGAT is
included as a representative neural KG-aware baseline and is evaluated under the same data split, validation procedure, and all-items ranking protocol. We report new KGAT results rather than reusing originally published numbers, since split or sampled-evaluation differences would not be comparable. iALS and BPR-MF are tuned over latent
dimensions and regularization, with an additional confidence-weight grid for
iALS, using a multi-threaded \texttt{implicit} solver; LightGCN and KGAT follow
their original hyperparameters. Stochastic baselines, including KGAT, are run
with three seeds; means are reported and all standard deviations are below
$0.001$.

\subsection{Implementation Details}
\label{sec:setup:impl}

All non-neural linear models and classical baselines run CPU-only on a single
Apple MacBook Air (M4, $32$ GB unified memory). KGAT is trained on an NVIDIA
A100 (40 GB) GPU because CPU training was impractical under our time budget; we
report the device explicitly in Sect.~\ref{sec:efficiency} and evaluate all
methods under the same train/validation/test split and all-items ranking
protocol. For TailSpec-EASE we tune the ridge weight $\lambda$, KG strength
$\mu$, tail exponent $\gamma$, and diffusion depth $K$ on validation, while
fixing neighborhood sizes and pruning thresholds across datasets. We report two
selection rules: \emph{best-overall}, maximizing validation NDCG@20, and
\emph{tail-constrained}, maximizing validation tail Recall@20 among
configurations within $1\%$ of the best validation NDCG@20.

\section{Main Results}
\label{sec:results}

\subsection{Overall and Long-Tail Accuracy}
\label{sec:results:main}

Table~\ref{tab:main} reports ranking quality and long-tail performance. We make
three observations. First, KGAT is strong in overall NDCG, confirming that the
comparison includes a competitive KG-aware neural model. TailSpec-EASE remains
close in overall accuracy: it is best on Amazon-book, nearly tied with KGAT on
Last-FM, competitive with KGAT and the global EASE$^\text{R}$ upper bound on
ML-1M, and trades a small amount of overall NDCG for stronger tail recall on
Yelp2018. Second, TailSpec-EASE consistently improves Tail Recall@20 over KGAT
and over its no-KG counterpart Local EASE. Against KGAT, the tail-constrained
TailSpec setting improves tail recall on all four datasets; against Local EASE,
all tail-recall gains are significant under paired bootstrap. Third, the baselines
occupy different operating points: matrix factorization and KG-aware neural
models can be strong in overall NDCG, while ItemKNN and BPR-MF may expose
tail items at the cost of large accuracy drops. TailSpec-EASE offers a more
balanced accuracy--tail--efficiency trade-off.

To confirm that the long-tail advantage over KGAT is robust, we apply the same
paired user-level bootstrap to the Tail Recall@20 difference between the
tail-constrained TailSpec-EASE and KGAT. The difference is positive on all four
datasets and significant on Amazon-book, Last-FM, and Yelp2018; only on ML-1M,
where the absolute difference is small, is it not significant. This supports the conclusion that TailSpec-EASE provides stronger tail exposure
than KGAT, rather than benefiting only from comparison with its no-KG backbone.

\begin{table}[t]
\centering
\caption{Overall accuracy (NDCG@20) and long-tail performance (Tail Recall@20)
on the test sets. KGAT$^{*,\mathrm{G}}$ is a representative KG-aware neural
baseline trained on GPU and evaluated under the same train/validation/test split
and all-items ranking protocol. For stochastic baselines (marked $*$), means
over three seeds are shown; all standard deviations are below $0.001$ and omitted
for compactness. TailSpec results are highlighted in \textbf{bold}. $\dagger$ marks
TailSpec gains over Local EASE that are significant under a paired user-level
bootstrap ($95\%$ CI excluding zero). OOM denotes out-of-memory at this catalog
size (Sect.~\ref{sec:efficiency}).}
\label{tab:main}
\setlength{\tabcolsep}{3pt}
\resizebox{\textwidth}{!}{%
\begin{tabular}{lcccccccc}
\toprule
& \multicolumn{2}{c}{ML-1M} & \multicolumn{2}{c}{Amazon-book}
& \multicolumn{2}{c}{Last-FM} & \multicolumn{2}{c}{Yelp2018}\\
\cmidrule(lr){2-3}\cmidrule(lr){4-5}\cmidrule(lr){6-7}\cmidrule(lr){8-9}
Method & NDCG & T-Rec & NDCG & T-Rec & NDCG & T-Rec & NDCG & T-Rec\\
\midrule
Popularity   & .1627 & .0000 & .0124 & .0000 & .0106 & .0000 & .0109 & .0000\\
ItemKNN      & .2629 & .0046 & .0774 & .0563 & .0653 & .0487 & .0328 & .0236\\
BPR-MF$^{*}$ & .1327 & .0744 & .0458 & .0214 & .0601 & .0208 & .0198 & .0008\\
iALS$^{*}$   & .2573 & .0006 & .0897 & .0095 & .0865 & .0219 & .0477 & .0000\\
LightGCN$^{*}$ & .2809 & .0084 & .0699 & .0119 & .0672 & .0242 & .0440 & .0006\\
KGAT$^{*,\mathrm{G}}$ & .3002 & .0067 & .0986 & .0453 & .0927 & .0820 & .0464 & .0049\\
global EASE$^\text{R}$ & .3025 & .0002 & .0954 & .0247 & OOM & OOM & OOM & OOM\\
Local EASE   & .2969 & .0023 & .0953 & .0329 & .0747 & .0413 & .0442 & .0030\\
\midrule
TailSpec (Ours, overall) & \textbf{.2985} & \textbf{.0038} & \textbf{.0994} & \textbf{.0553} & \textbf{.0914} & \textbf{.0897} & \textbf{.0450} & \textbf{.0046}\\
TailSpec (Ours, tail)    & \textbf{.2976} & \textbf{.0084}$^{\dagger}$ & \textbf{.0994} & \textbf{.0553}$^{\dagger}$ & \textbf{.0926} & \textbf{.0958}$^{\dagger}$ & \textbf{.0446} & \textbf{.0071}$^{\dagger}$\\
\bottomrule
\end{tabular}%
}
\end{table}

\subsection{Statistical Significance}
\label{sec:results:significance}
Table~\ref{tab:bootstrap} reports paired user-level bootstrap comparisons between
TailSpec-EASE and its no-KG counterpart Local EASE. The tail-metric differences
are positive and significant on all four datasets, confirming that the long-tail
improvements are not driven by a few users. Overall NDCG also improves
significantly on Amazon-book, Last-FM, and Yelp2018, while the ML-1M overall
difference is small and not significant. This is consistent with the goal of using
the KG prior mainly to improve tail exposure without harming head-item
accuracy.

\begin{table}[t]
\centering
\caption{Paired user-level bootstrap (1,000 resamples) of the per-user metric
difference TailSpec-EASE minus Local EASE on the test set. Each cell reports the
mean difference; $\dagger$ indicates that the 95\% bootstrap confidence interval
excludes zero, and n.s. denotes not significant. Confidence intervals are omitted
for compactness.}
\label{tab:bootstrap}
\setlength{\tabcolsep}{6pt}
\renewcommand{\arraystretch}{0.95}
\begin{tabular}{lcccc}
\toprule
Metric & ML-1M & Amazon-book & Last-FM & Yelp2018\\
\midrule
NDCG@20
& $+.0006$ {\footnotesize n.s.}
& $+.0041^{\dagger}$
& $+.0179^{\dagger}$
& $+.0004^{\dagger}$\\
Recall@20
& $-.0004$ {\footnotesize n.s.}
& $+.0091^{\dagger}$
& $+.0194^{\dagger}$
& $+.0008^{\dagger}$\\
Tail NDCG@20
& $+.0031^{\dagger}$
& $+.0116^{\dagger}$
& $+.0347^{\dagger}$
& $+.0021^{\dagger}$\\
Tail Recall@20
& $+.0061^{\dagger}$
& $+.0224^{\dagger}$
& $+.0544^{\dagger}$
& $+.0041^{\dagger}$\\
\bottomrule
\end{tabular}
\end{table}

\section{Model-Level versus Score-Level Knowledge Injection}
\label{sec:postproc}

A natural concern is whether the KG gains could be obtained by post-processing
Local EASE scores instead of modifying the learning objective. We therefore
compare TailSpec-EASE with two score-level baselines that use the same KG
operators: Smooth applies the relation-aware operator $\mathbf{P}_{\boldsymbol{\beta}}$,
and Diffuse applies the diffusion prior $\mathbf{H}_K$, each with a
validation-tuned mixing weight. The KG signal and Local EASE backbone are held
fixed; only the point of injection differs.

Table~\ref{tab:postproc} shows that model-level injection is consistently more
effective. TailSpec-EASE obtains higher NDCG@20 and higher Tail Recall@20 than
both score-level baselines on all four datasets. The difference is clearest on
Yelp2018, where score-level smoothing barely moves tail recall and slightly
reduces NDCG, while TailSpec-EASE improves both. Last-FM is the closest case:
Diffuse is competitive, but still does not match the model-level result.

\begin{table}[t]
\centering
\caption{Model-level (TailSpec) versus score-level KG injection (Smooth,
Diffuse), all using the same KG operators on the same Local EASE backbone. Test
NDCG@20 and Tail Recall@20; best per group in \textbf{bold}. Score-level
methods are tuned on validation; we report their tail-constrained selection.}
\label{tab:postproc}
\setlength{\tabcolsep}{4pt}
\resizebox{\textwidth}{!}{%
\begin{tabular}{lcccccccc}
\toprule
& \multicolumn{2}{c}{ML-1M} & \multicolumn{2}{c}{Amazon-book}
& \multicolumn{2}{c}{Last-FM} & \multicolumn{2}{c}{Yelp2018}\\
\cmidrule(lr){2-3}\cmidrule(lr){4-5}\cmidrule(lr){6-7}\cmidrule(lr){8-9}
Method & NDCG & T-Rec & NDCG & T-Rec & NDCG & T-Rec & NDCG & T-Rec\\
\midrule
Local EASE (no KG)  & .2969 & .0023 & .0953 & .0329 & .0747 & .0413 & .0442 & .0030\\
Smooth (score-level)  & .2966 & .0024 & .0973 & .0378 & .0896 & .0726 & .0440 & .0036\\
Diffuse (score-level) & .2969 & .0024 & .0991 & .0480 & .0901 & .0741 & .0439 & .0034\\
TailSpec (Ours, model-level) & \textbf{.2976} & \textbf{.0084} & \textbf{.0994} & \textbf{.0553} & \textbf{.0926} & \textbf{.0958} & \textbf{.0446} & \textbf{.0071}\\
\bottomrule
\end{tabular}%
}
\end{table}

\begin{figure}[t]
\centering
\includegraphics[width=0.70\textwidth]{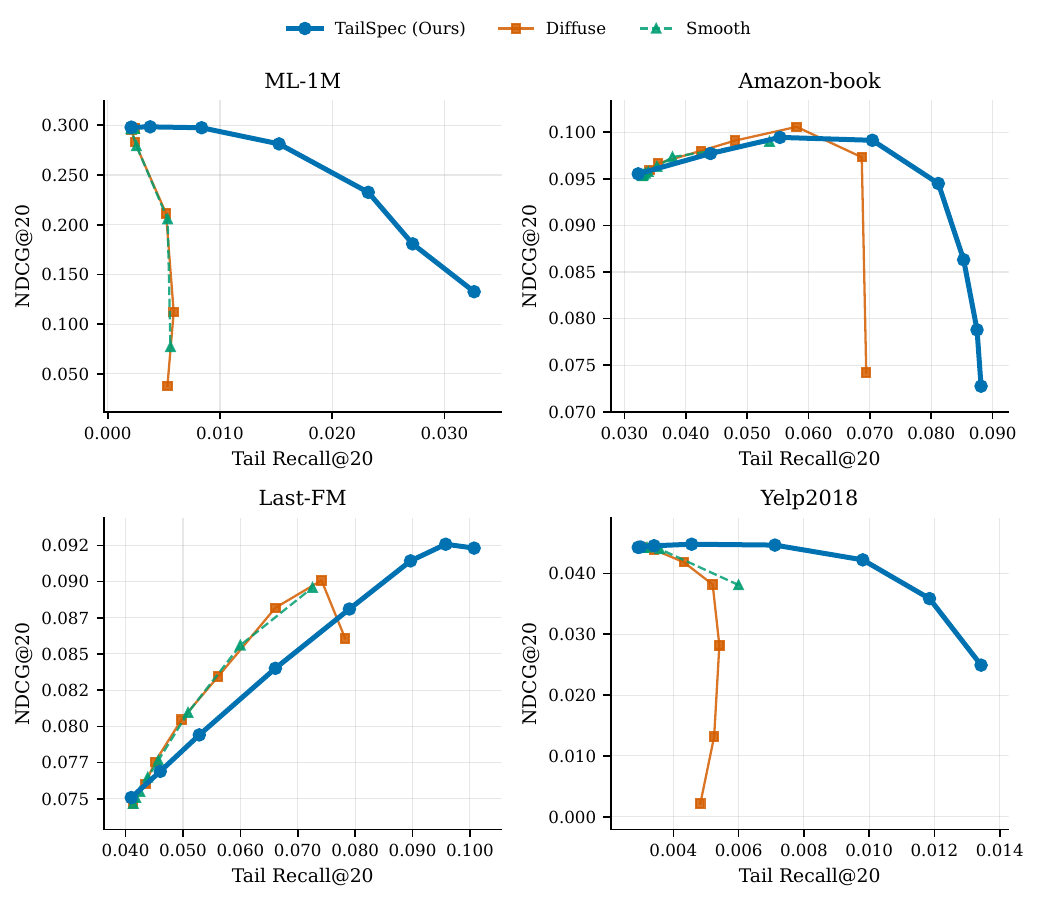}
\caption{Accuracy--tail trade-off frontiers on the (Tail Recall@20, NDCG@20)
plane; axes scaled per dataset, upper-right is better. TailSpec-EASE sweeps the
KG strength $\mu$; Smooth and Diffuse sweep the score-level mixing weight.
Across datasets TailSpec-EASE reaches tail recall unattainable by score-level
post-processing without collapsing accuracy; the post-processing baselines lose
accuracy sharply at large weights, most visibly on ML-1M and Yelp2018, and are
competitive only on the KG-dense Last-FM.}
\label{fig:frontier}
\end{figure}

Figure~\ref{fig:frontier} traces the full accuracy--tail frontier by sweeping the
KG strength $\mu$ for TailSpec-EASE and the score-level mixing weight for
Smooth and Diffuse. On KG-dense Last-FM, score-level diffusion is competitive;
on ML-1M and Yelp2018, however, the post-processing curves collapse once their
weight grows. Model-level injection is more reliable because the KG prior
participates in fitting the reconstruction weights: the data term can override the
prior for head items, while the popularity-adaptive strength $\mu g_i$ lets the
prior dominate for tail items. Score-level propagation instead applies a single
global mixing weight after training and cannot modulate KG influence per item.

\section{Ablation and Sensitivity}
\label{sec:ablation}

We isolate three design choices: the tail-adaptive gate, relation typing, and
diffusion depth. Unless otherwise stated, sensitivity analyses are performed on
validation, leaving model selection independent of the test set.

\paragraph{Tail-adaptive gate.}
The gate exponent $\gamma$ in Eq.~\eqref{eq:gate} shapes the accuracy--tail
frontier rather than acting as an independent source of gain. With $\gamma=0$,
the prior is applied uniformly and often reaches the best overall operating
point. Positive $\gamma$ protects head items when $\mu$ is large, allowing more
aggressive tail exposure without collapsing head accuracy. The two selection
rules in Table~\ref{tab:main} reflect this trade-off.

\paragraph{Relation typing.}
An untyped variant that pours all triples into a single graph confirms that
relation typing mainly benefits the tail. Across all four datasets, typed
relations improve Tail Recall@20, although their effect on overall NDCG is
dataset-dependent. The gains are clearest on ML-1M and Amazon-book: on ML-1M,
tail recall rises from $.0030$ to $.0084$, and on Amazon-book both NDCG and
tail recall improve. On Last-FM the typed and untyped variants are nearly
identical, while on Yelp2018 typing trades a small amount of NDCG for higher
tail recall. This supports using typed relation graphs without making relation
typing the main source of the overall gain.

\paragraph{Diffusion depth.}
We vary the diffusion depth $K$ in Eq.~\eqref{eq:ppr} with other
hyperparameters fixed. Here $K=0$ is a no-propagation anchor: after the identity
self-entry is removed in Eq.~\eqref{eq:hi}, the KG prior contributes no
propagated semantic neighbors. Table~\ref{tab:depth} shows that one hop captures
essentially the full benefit. Moving from $K=0$ to $K=1$ sharply improves tail
recall, especially on Last-FM, while $K=2$ and $K=3$ provide no consistent gain.
Thus the useful KG signal lies mainly in direct semantic neighbors, and the
default $K=1$ is both accurate and cheap.

\begin{table}[t]
\centering
\caption{Diffusion-depth sensitivity on validation (NDCG@20 / Tail Recall@20), with
all other hyperparameters fixed. $K=0$ removes propagated KG neighbors after
self-entry removal and serves as the no-propagation anchor. A single hop captures the
full benefit; deeper propagation does not help.}
\label{tab:depth}
\setlength{\tabcolsep}{5pt}
\renewcommand{\arraystretch}{0.95}
\resizebox{0.9\textwidth}{!}{%
\begin{tabular}{lcccccccc}
\toprule
& \multicolumn{2}{c}{ML-1M} & \multicolumn{2}{c}{Amazon-book}
& \multicolumn{2}{c}{Last-FM} & \multicolumn{2}{c}{Yelp2018}\\
\cmidrule(lr){2-3}\cmidrule(lr){4-5}\cmidrule(lr){6-7}\cmidrule(lr){8-9}
Depth & NDCG & T-Rec & NDCG & T-Rec & NDCG & T-Rec & NDCG & T-Rec\\
\midrule
$K=0$ & .2014 & .0011 & .1424 & .0523 & .1130 & .0223 & .0540 & .0022\\
$K=1$ & .2009 & \textbf{.0075} & \textbf{.1461} & \textbf{.0906} & \textbf{.1902} & \textbf{.2509} & \textbf{.0544} & \textbf{.0135}\\
$K=2$ & \textbf{.2017} & .0064 & \textbf{.1462} & .0902 & .1893 & .2480 & .0543 & .0121\\
$K=3$ & .2018 & .0060 & .1462 & .0902 & .1890 & .2474 & .0543 & .0121\\
\bottomrule
\end{tabular}%
}
\end{table}

\section{Efficiency and Scalability}
\label{sec:efficiency}

A central design goal of TailSpec-EASE is CPU-only scalability. Table~\ref{tab:efficiency}
reports a representative Amazon-book timing study together with accuracy and
tail recall. TailSpec-EASE trains in $37$ seconds on CPU, only modestly slower
than Local EASE, because the KG prior affects small local systems rather than a
dense global solve. It also compares favorably with neural baselines. LightGCN
requires $15{,}800$ seconds on CPU and obtains lower NDCG@20 and Tail
Recall@20. KGAT is a stronger KG-aware neural baseline, obtaining NDCG@20 of
$0.0986$ and Tail Recall@20 of $0.0453$, but requires $2{,}584$ seconds on GPU.
TailSpec-EASE achieves higher NDCG@20 ($0.0994$) and Tail Recall@20 ($0.0553$)
in $37$ seconds on CPU.

\begin{table}[t]
\centering
\caption{Representative Amazon-book efficiency study. Times exclude evaluation.
KGAT runs on an NVIDIA A100 GPU, while the other methods run CPU-only on
the M4 machine. The comparison is not hardware-identical, but reflects practical
training and hardware cost.}
\label{tab:efficiency}
\setlength{\tabcolsep}{4pt}
\renewcommand{\arraystretch}{0.95}
\resizebox{\textwidth}{!}{%
\begin{tabular}{lcccc}
\toprule
Method & KG used & Device & Training time & NDCG@20 / T-Rec@20\\
\midrule
Local EASE & No & CPU & 12\,s & .0953 / .0329\\
global EASE$^\text{R}$ & No & CPU & 182\,s & .0954 / .0247\\
LightGCN & No & CPU & 15{,}800\,s ($\approx$4.4\,h) & .0699 / .0119\\
KGAT & Yes & GPU & 2{,}584\,s ($\approx$43.1\,min) & .0986 / .0453\\
\textbf{TailSpec-EASE (Ours)} & Yes & CPU & \textbf{37\,s} & \textbf{.0994 / .0553}\\
\bottomrule
\end{tabular}%
}
\end{table}

Because KGAT is trained on GPU while TailSpec-EASE is trained on CPU, this
comparison should not be read as a hardware-identical speedup. Instead, it
highlights the practical deployment difference: TailSpec-EASE obtains a
competitive, and on Amazon-book stronger, accuracy--tail trade-off without GPU
training. The comparison with global EASE$^\text{R}$ further shows why locality
matters: on the larger Last-FM and Yelp2018 catalogs, the global dense inversion
exceeds the $32$\,GB memory budget, whereas TailSpec-EASE remains feasible
because its memory footprint is governed by the local neighborhood size rather
than the catalog size.

\section{Discussion and Conclusion}
\label{sec:discussion}

TailSpec-EASE is most useful when the item-side KG aligns with interaction
structure. The gains are largest on KG-dense Last-FM and smaller on Yelp2018,
where the KG is comparatively sparse relative to the catalog. This is a condition
of applicability rather than a failure mode: when the KG is useful, model-level
injection yields large long-tail gains; when it is weak, the model degrades
gracefully toward the Local EASE backbone. The method also makes the head--tail
trade-off explicit. The best-overall selection rule favors NDCG@20, while the
tail-constrained rule selects a point with stronger long-tail exposure. This
matters on Yelp2018, where iALS attains slightly higher overall NDCG but
recommends almost no tail items. TailSpec-EASE deliberately occupies a different
operating point, trading a small amount of overall accuracy for statistically
significant long-tail gains.

The diffusion-depth study shows that TailSpec-EASE does not need deep multi-hop
KG propagation. A single hop captures essentially the full benefit, suggesting
that the useful KG signal lies mainly in direct semantic neighbors. In summary,
TailSpec-EASE brings item-side KG structure into the efficient regime of
closed-form item-based recommendation. By combining a relation-aware spectral
prior, local reconstruction, and popularity-adaptive regularization, it improves
long-tail recommendation while remaining CPU-efficient and scalable to catalogs
where global closed-form models are infeasible. Future work includes learning
relation weights instead of selecting them on validation, and extending the local
formulation to user-side knowledge, temporal recommendation, or session-based
settings.

\begin{credits}
\subsubsection{\discintname}
The authors have no competing interests to declare that are relevant to the
content of this article.
\end{credits}

%
%
% ---- Bibliography ----
%
\bibliographystyle{splncs04}
\bibliography{references}

@inproceedings{wang2019kgat,
  author    = {Wang, X. and He, X. and Cao, Y. and Liu, M. and Chua, T. S.},
  title     = {{KGAT}: Knowledge Graph Attention Network for Recommendation},
  booktitle = {Proceedings of the 25th ACM SIGKDD International Conference on Knowledge Discovery and Data Mining (KDD)},
  pages     = {950--958},
  year      = {2019},
  doi       = {10.1145/3292500.3330989}
}

@inproceedings{wang2018ripplenet,
  author    = {Wang, H. and Zhang, F. and Wang, J. and Zhao, M. and Li, W. and Xie, X. and Guo, M.},
  title     = {{RippleNet}: Propagating User Preferences on the Knowledge Graph for Recommender Systems},
  booktitle = {Proceedings of the 27th ACM International Conference on Information and Knowledge Management (CIKM)},
  pages     = {417--426},
  year      = {2018},
  doi       = {10.1145/3269206.3271739}
}

@inproceedings{wang2019kgcn,
  author    = {Wang, H. and Zhao, M. and Xie, X. and Li, W. and Guo, M.},
  title     = {Knowledge Graph Convolutional Networks for Recommender Systems},
  booktitle = {Proceedings of the World Wide Web Conference (WWW)},
  pages     = {3307--3313},
  year      = {2019},
  doi       = {10.1145/3308558.3313417}
}

@inproceedings{wang2019kgnnls,
  author    = {Wang, H. and Zhang, F. and Zhang, M. and Leskovec, J. and Zhao, M. and Li, W. and Wang, Z.},
  title     = {Knowledge-aware Graph Neural Networks with Label Smoothness Regularization for Recommender Systems},
  booktitle = {Proceedings of the 25th ACM SIGKDD International Conference on Knowledge Discovery and Data Mining (KDD)},
  pages     = {968--977},
  year      = {2019},
  doi       = {10.1145/3292500.3330836}
}

@inproceedings{ning2011slim,
  author    = {Ning, X. and Karypis, G.},
  title     = {{SLIM}: Sparse Linear Methods for Top-{N} Recommender Systems},
  booktitle = {Proceedings of the IEEE International Conference on Data Mining (ICDM)},
  pages     = {497--506},
  year      = {2011},
  doi       = {10.1109/ICDM.2011.134}
}

@inproceedings{steck2019ease,
  author    = {Steck, H.},
  title     = {Embarrassingly Shallow Autoencoders for Sparse Data},
  booktitle = {Proceedings of the World Wide Web Conference (WWW)},
  pages     = {3251--3257},
  year      = {2019},
  doi       = {10.1145/3308558.3313710}
}

@inproceedings{dacrema2019recsys,
  author    = {Ferrari Dacrema, M. and Cremonesi, P. and Jannach, D.},
  title     = {Are We Really Making Much Progress? A Worrying Analysis of Recent Neural Recommendation Approaches},
  booktitle = {Proceedings of the 13th ACM Conference on Recommender Systems (RecSys)},
  pages     = {101--109},
  year      = {2019},
  doi       = {10.1145/3298689.3347058}
}

@inproceedings{hu2008ials,
  author    = {Hu, Y. and Koren, Y. and Volinsky, C.},
  title     = {Collaborative Filtering for Implicit Feedback Datasets},
  booktitle = {Proceedings of the IEEE International Conference on Data Mining (ICDM)},
  pages     = {263--272},
  year      = {2008},
  doi       = {10.1109/ICDM.2008.22}
}

@inproceedings{rendle2009bpr,
  author    = {Rendle, S. and Freudenthaler, C. and Gantner, Z. and Schmidt-Thieme, L.},
  title     = {{BPR}: Bayesian Personalized Ranking from Implicit Feedback},
  booktitle = {Proceedings of the 25th Conference on Uncertainty in Artificial Intelligence (UAI)},
  pages     = {452--461},
  year      = {2009}
}

@inproceedings{sarwar2001itemknn,
  author    = {Sarwar, B. and Karypis, G. and Konstan, J. and Riedl, J.},
  title     = {Item-based Collaborative Filtering Recommendation Algorithms},
  booktitle = {Proceedings of the 10th International Conference on World Wide Web (WWW)},
  pages     = {285--295},
  year      = {2001},
  doi       = {10.1145/371920.372071}
}

@inproceedings{he2020lightgcn,
  author    = {He, X. and Deng, K. and Wang, X. and Li, Y. and Zhang, Y. and Wang, M.},
  title     = {{LightGCN}: Simplifying and Powering Graph Convolution Network for Recommendation},
  booktitle = {Proceedings of the 43rd International ACM SIGIR Conference on Research and Development in Information Retrieval (SIGIR)},
  pages     = {639--648},
  year      = {2020},
  doi       = {10.1145/3397271.3401063}
}

@inproceedings{park2008longtail,
  author    = {Park, Y. J. and Tuzhilin, A.},
  title     = {The Long Tail of Recommender Systems and How to Leverage It},
  booktitle = {Proceedings of the ACM Conference on Recommender Systems (RecSys)},
  pages     = {11--18},
  year      = {2008},
  doi       = {10.1145/1454008.1454012}
}

@inproceedings{steck2011popularity,
  author    = {Steck, H.},
  title     = {Item Popularity and Recommendation Accuracy},
  booktitle = {Proceedings of the ACM Conference on Recommender Systems (RecSys)},
  pages     = {125--132},
  year      = {2011},
  doi       = {10.1145/2043932.2043957}
}

@inproceedings{abdollahpouri2017controlling,
  author    = {Abdollahpouri, H. and Burke, R. and Mobasher, B.},
  title     = {Controlling Popularity Bias in Learning-to-Rank Recommendation},
  booktitle = {Proceedings of the 11th ACM Conference on Recommender Systems (RecSys)},
  pages     = {42--46},
  year      = {2017},
  doi       = {10.1145/3109859.3109912}
}

@inproceedings{steck2018calibrated,
  author    = {Steck, H.},
  title     = {Calibrated Recommendations},
  booktitle = {Proceedings of the 12th ACM Conference on Recommender Systems (RecSys)},
  pages     = {154--162},
  year      = {2018},
  doi       = {10.1145/3240323.3240372}
}

@inproceedings{liu2020longtail,
  author    = {Liu, S. and Zheng, Y.},
  title     = {Long-tail Session-based Recommendation},
  booktitle = {Proceedings of the 14th ACM Conference on Recommender Systems (RecSys)},
  pages     = {509--514},
  year      = {2020},
  doi       = {10.1145/3383313.3412222}
}

@book{chung1997spectral,
  author    = {Chung, F. R. K.},
  title     = {Spectral Graph Theory},
  publisher = {American Mathematical Society},
  year      = {1997}
}

@techreport{page1999pagerank,
  author      = {Page, L. and Brin, S. and Motwani, R. and Winograd, T.},
  title       = {The {PageRank} Citation Ranking: Bringing Order to the Web},
  institution = {Stanford InfoLab},
  year        = {1999}
}

@inproceedings{gori2007itemrank,
  author    = {Gori, M. and Pucci, A.},
  title     = {{ItemRank}: A Random-Walk Based Scoring Algorithm for Recommender Engines},
  booktitle = {Proceedings of the 20th International Joint Conference on Artificial Intelligence (IJCAI)},
  pages     = {2766--2771},
  year      = {2007}
}

@article{jarvelin2002ndcg,
  author  = {J{\"a}rvelin, K. and Kek{\"a}l{\"a}inen, J.},
  title   = {Cumulated Gain-based Evaluation of {IR} Techniques},
  journal = {ACM Transactions on Information Systems},
  volume  = {20},
  number  = {4},
  pages   = {422--446},
  year    = {2002},
  doi     = {10.1145/582415.582418}
}

@article{harper2015movielens,
  author    = {Harper, F. Maxwell and Konstan, Joseph A.},
  title     = {The {MovieLens} Datasets: History and Context},
  journal   = {ACM Transactions on Interactive Intelligent Systems},
  volume    = {5},
  number    = {4},
  pages     = {19:1--19:19},
  year      = {2015},
  doi       = {10.1145/2827872}
}

@inproceedings{mcauley2015image,
  author    = {McAuley, Julian and Targett, Christopher and Shi, Qinfeng and van den Hengel, Anton},
  title     = {Image-Based Recommendations on Styles and Substitutes},
  booktitle = {Proceedings of the 38th International ACM SIGIR Conference on Research and Development in Information Retrieval (SIGIR)},
  pages     = {43--52},
  year      = {2015},
  doi       = {10.1145/2766462.2767755}
}

@inproceedings{cantador2011hetrec,
  author    = {Cantador, Iv{\'a}n and Brusilovsky, Peter and Kuflik, Tsvi},
  title     = {Second Workshop on Information Heterogeneity and Fusion in Recommender Systems ({HetRec2011})},
  booktitle = {Proceedings of the 5th ACM Conference on Recommender Systems (RecSys)},
  pages     = {387--388},
  year      = {2011},
  doi       = {10.1145/2043932.2044016}
}

@misc{yelp2018dataset,
  author    = {{Yelp}},
  title     = {Yelp Open Dataset},
  howpublished = {\url{https://www.yelp.com/dataset}},
  year      = {2018},
  note      = {Accessed: 2026-06-10}
}

\end{document}